\documentclass[11pt,twoside]{article}

\usepackage{asp2006}
\usepackage{epsfig}

\def\araa{ARAA}
\def\mnras{MNRAS}
\def\apj{APJ}

\def\apjl{APJL}
\def\aap{AAP}

\newcommand{\Wt}{\tilde{W}}
\newcommand{\vt}{{\vec \theta}}
\newcommand{\vU}{{\vec{ U}}}
\def\V{{\cal V}}
\def\N{{\cal N}}
\def\HI{{\rm HI}}

\begin{document}
\title{Probing Turbulence in the Interstellar Medium of Galaxies} 
\author{Prasun Dutta$^1$, Ayesha Begum$^2$, Somnath Bharadwaj$^3$ and Jayaram
  N. Chengalur$^4$}
\affil{$^{1,3}$ Center for Theoretical Studies, IIT-Kharagpur, India\\
$^{2}$Department of Astronomy, University of Wisconsin\\
$^{4}$NCRA--TIFR Pune, India}

\begin{abstract} 
The power spectrum 
of HI intensity fluctuation of the interstellar medium carries
information of the turbulence dynamics therein. We present a method to
estimate the power spectrum of HI intensity 
fluctuation using radio interferometric observations. The method
involves correlating the visibilities in the $u-v$ plane at different
baselines. This method is particularly use full for evaluating the
power spectrum for the faint dwarf galaxies. We apply this method to 3
spiral galaxies and 5 dwarf galaxies. The measured power
spectrum seem to follow a power law $P_{\rm HI}(U) = A
U^{\alpha}$, suggesting turbulence to be operational. Further,
depending on the slope of 
the power spectrum, we expect the presence of 2D and 3D turbulence
in those galaxies. 
\end{abstract}


\section{Introduction}   
Evidence has been mounting in recent years that turbulence plays an important
role in the physics of the ISM as well as in governing star formation. It is
believed that turbulence is responsible for generating the hierarchy
of structures present across  a range of spatial scales in the ISM
 (e.g. \citealt{ES04I}; \citealt{ES04II}). In such 
models the ISM has a fractal structure and the power spectrum of  intensity
fluctuations is a power law, indicating that there is no preferred
``cloud" size. 

On the observational front, power spectrum analysis of  HI
intensity fluctuations is  an important technique to probe the
structure of the neutral ISM in galaxies.   
The power spectra of the HI intensity fluctuations in our own galaxy, the LMC   
and the SMC all show power  law behaviour (\citealt{CD83}; \citealt{GR93};
\citealt{DD00};  \citealt{EK01}; \citealt{SSD99}) which is a
characteristic of a turbulent medium.

Recently \citet{AJS06} have presented a visibility based formalism for
determining the power spectrum  of  HI intensity fluctuations in
galaxies with extremely weak emission.  This
formalism was applied to a dwarf
galaxy, DDO~210 and a spiral galaxy NGC~628
\citep{DBBC08}. Interestingly, the HI power spectrum of both of these
galaxies  were found to be power law. In this report we present a
elaborate description of this method. We also report the result of
estimation of the power spectrum of 3 spiral galaxies and 5 dwarf galaxies.

\section{ A visibility based power spectrum estimator}

The specific intensity of the HI emission from a  galaxy 
may be modelled as 
\begin{equation}
I_{\nu}(\vt) \ =\ W_{\nu}(\vt)\left[ \bar{I}_{\nu} \ +\ \delta
  I_{\nu}(\vt) \right ]\,.
\end{equation}
Here we have assumed that this is the sum of a  smooth component and a 
fluctuating component.
We express the fluctuating component of the specific intensity as 
$W_\nu(\vt) \, \delta I_\nu(\vt)$, where $\delta   I_\nu(\vt)$ is
assumed to be a statistically homogeneous and isotropic
stochastic fluctuation and  $W_\nu(\vt)$ is the window function 
which quantify the overall large scale HI distribution of the galaxy.
Since, the angular extent of the galaxies that we consider here is
much smaller than the primary beam, we can write the visibility, the
quantity directly measured by the radio interferometers, as  
\begin{equation}
\V_\nu(\vU)=\Wt(\vU) \bar{I}_\nu + \Wt(\vU)  \otimes
\tilde{\delta I_{\nu}}(\vU)  + \N_\nu(\vU)
\label{eq:vis2}
\end{equation}
where  the tilde $\tilde{\,}$ denotes the Fourier transform of the
corresponding quantity and $\otimes$ denotes a convolution. 
In addition to the signal,  each visibility
also contains a system noise contribution $\N_\nu(\vU)$ which we have
introduced in  
eq. (\ref{eq:vis2}). The noise in each visibility is a Gaussian random
variable and the noise in the visibilities at two different baselines
$\vU$ and $\vU^{'}$ is uncorrelated. 
 At larger baselines, where the effect of the window function can be
 neglected, we can write,
\begin{equation}
\V_\nu(\vU)= \Wt(\vU)  \otimes
\tilde{\delta I_{\nu}}(\vU)  + \N_\nu(\vU)
\label{eq:vis3}
\end{equation}
 We use the power spectrum of HI intensity fluctuations $P_{\HI}(U)$
defined as
\begin{equation}
\langle \tilde{\delta I_\nu}(\vU) \tilde{\delta I_\nu}^{*}(\vU')
\rangle =\delta^2(\vU-\vU') \, P_{\HI}(U)
\end{equation}
to quantify the statistical properties of the intensity
fluctuations. We use angular averaging in place of 
the ensemble averaging denoted by the angular brackets.
The square of the visibilities can, in principle, be used to estimate
$P_{\HI}(U)$ 
\begin{equation}
\langle\ \V_{\nu}(\vU)\V^{*}_{\nu}(\vU)\ \rangle \ = \left
|\Wt_{\nu}(\vU)  \right|^{2} \otimes   \ P_{HI}(\vU) + \langle \mid
\N_\nu (\vU) \mid^2 \rangle
\end{equation}
The last term  $\langle \mid\N_\nu (\vU) \mid^2 \rangle$, which  is
the noise variance, introduces a positive bias in estimating the power
spectrum.  The  noise bias can be orders of magnitude larger than the
power spectrum for the  faint external galaxies considered here.
The  problem of noise bias can be avoided by correlating  visibilities
at two different baselines for which  the noise is expected to be
uncorrelated. We define the power spectrum estimator 
\begin{eqnarray}
\hat{\rm P}_{\rm HI}(\vU, \Delta \vU) &=&
\langle\ \V_{\nu}(\vU)\V^{*}_{\nu}(\vU+\Delta \vU)\ \rangle 
\nonumber  \\
&=& \int d^2 U' \, W_{\nu}(\vU-\vU') \, W^*_{\nu}(\vU+\Delta \vU-\vU')
\      P_{\HI}(\vU')  
\label{eq:corr}
\end{eqnarray}

In our analysis we  consider two different models for the window
function of a galaxy. We present them here in the Fourier transformed form. 
\begin{equation}
\tilde{W}_\nu(\vU)  =  2 ~\frac{J_{1}(2 \pi \theta_0 U)}
{2 \pi \theta_0 U}
\end{equation}
and 
\begin{equation}
\tilde{W}_\nu(\vU)  = \exp(- \pi^2 \theta_0^2 U^2/2)
\label{eq:gausw}
\end{equation}
  It is to note that the visibilities at two  different 
baselines  will be correlated only if  $|\Delta \vU| <
(\pi \theta_{0})^{-1}$, and not beyond. In  our
analysis we restrict the 
difference in baselines to  $|\Delta \vU| \ll  (\pi \theta_{0})^{-1}$
so that  
$\tilde W_\nu(\vU + \Delta \vU-\vU^{'}) \approx \tilde
W_\nu(\vU-\vU^{'})$ and   the estimator $\hat{\rm P}_{\rm
  HI}(\vU, \Delta \vU)$ no  longer depends on $\Delta \vU$. We then
use   the visibility correlation estimator 
\begin{eqnarray}
\hat{\rm P}_{\rm HI}(\vU) &=&
\langle\ \V_{\nu}(\vU)\V^{*}_{\nu}(\vU+\Delta \vU)\ \rangle 
\nonumber  \\
&=& \int d^2 U' \, \mid W_{\nu}(\vU-\vU') \mid^2 \,
P_{\HI}(\vU')  \,.
\label{eq:corrn}
\end{eqnarray}
We use the real part of the estimator $\hat{\rm P}_{\HI}(\vU)$ to estimate 
power spectrum $P_{\HI}(U)$.
 The real part is the power spectrum of
HI intensity fluctuations convolved with the square of the window
function. At large baselines, the effect of the window function can be
neglected and we then have  
\begin{equation}
\hat{\rm P}_{\rm HI}(\vU)= C \,  P_{\HI} (\vU) 
\label{eq:corra}
\end{equation}
where $C=\int  \mid \tilde W_\nu(U) \mid^2 ~d^2  U$ is a constant. 
 The estimator $\hat{\rm P}_{\HI}(\vU)$ also has a  small  imaginary
 part that arises  from the HI power spectrum  because  the
 assumption that     
$\tilde W_\nu(\vU + \Delta \vU - \vU^{'}) \approx \tilde W_\nu(\vU -
\vU^{'})$  is not strictly valid.  We use the requirement that the  
imaginary part of  $\hat{\rm P}_{\rm HI}(\vU)$ should be small 
 compared to the real part 
as a  self-consistency  check to determine the range  of validity of
our formalism.   

The  real and imaginary parts  of the 
measured value of the estimator $\hat{\rm P}_{\HI}(\vU)$ both have
uncertainties  arising from
 (1.) the sample variance and (2.) the system
noise.  We add both
these contributions to determine the $1-\sigma$ error-bars.  The
reader is referred to Section 3.2 of \citet{SS08} for further
details of the error estimation.  

\section{Result and Discussion}

 The  dwarf  galaxy data used here  is from  Giant Metrewave Radio
 Telescope (GMRT) observations.  
We use Very Large Array(VLA)  archival data, for the spiral galaxies
NGC~628, NGC~2403 and NGC~1058. Details can be found in \citet{DBBC09}
\begin{figure}
\begin{center}
\mbox{\epsfig{file=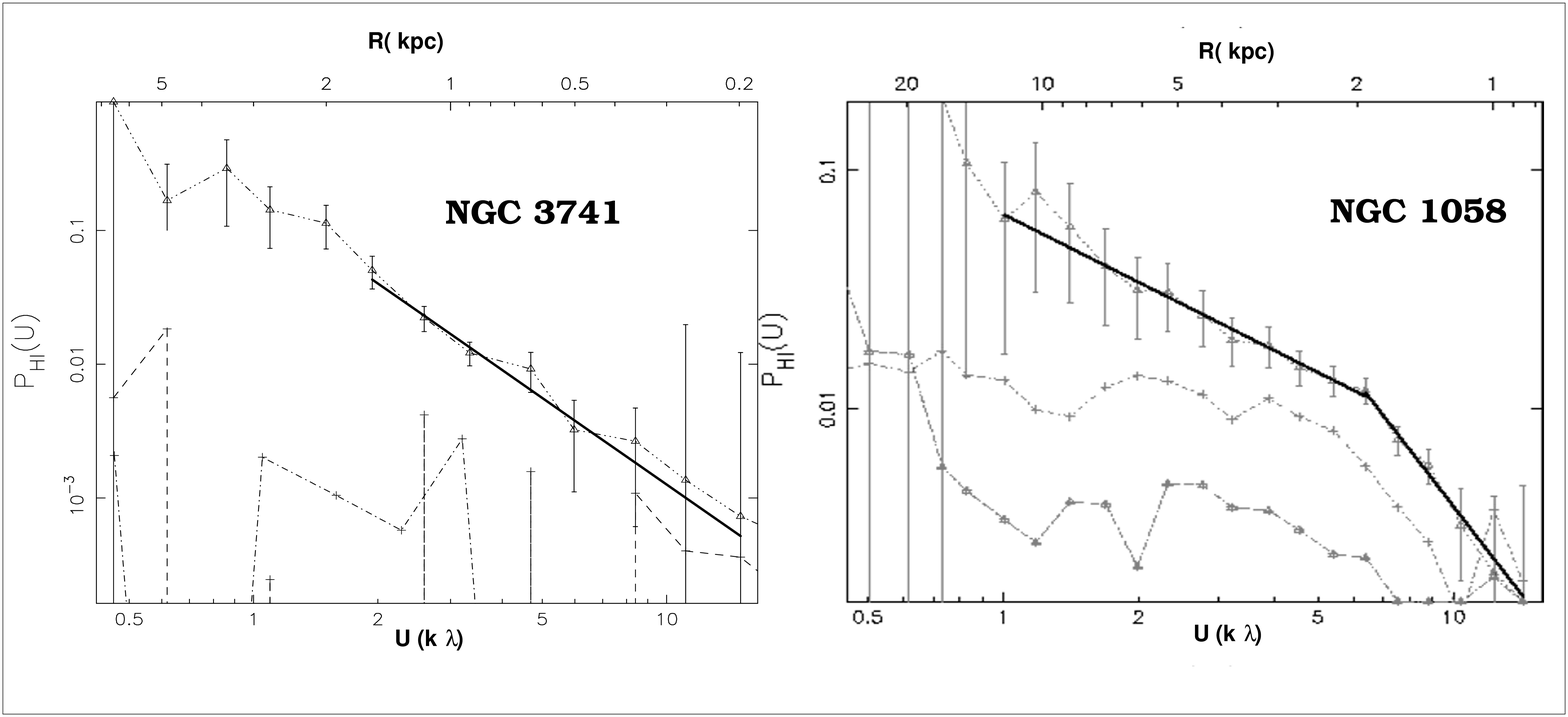,width=5.2in,height=2in,angle=0}}
\end{center}
\caption{Real part of the estimator, imaginary part of the estimator
   for the line channels and the real part of the estimator for the
   line-free channels are plotted from top to bottom respectively for
   the dwarf galaxy NGC~3741 (left) and the spiral galaxy NGC~1058
   (right). $1-\sigma$ error bars are also shown for the real part of
   the estimator for the line channels. Best fit is shown using a dark
   black line. Note that the best fit power spectrum for the galaxy
   NGC~1058 is a broken power law.
}
\label{fig:mom0}
\end{figure}
All the data are reduced in the usual way using  standard tasks in
classic AIPS \footnote{NRAO Astrophysical Image Processing System,   
a commonly used software for radio data processing.}. 
For each galaxy, after calibration the frequency channels with HI
emission were identified and a continuum image was  
made by combining the line free channels. The continuum was hence
subtracted from the data in the $u-v$ plane using  
the AIPS task UVSUB.  
We correlate the visibilities in each channel as in
eq. (\ref{eq:corrn}) and then average over channels to get $\hat{\rm
  P}_{\HI}(U)$. We then fit a power law 
 $A~U^{\alpha}$ to the estimator $\hat{\rm P}_{\HI}(U)$. The best fit  $A$
and $\alpha$ were determined through a $\chi^2$  
minimization. To test whether the impact of the window function  is
actually small we have convolved the best fit power spectrum with $\mid
\tilde{W}_{\nu}(U)\mid^2$. We estimate the goodness of
fit ($\chi^{2}$) to the data for the
convolved power spectrum.  The fit is accepted only after ensuring
that the effect of the convolution can actually be ignored. The
results are listed in table~1.

The galaxies in our sample have dichotomy in the slope $\alpha$
ranging from $-2.6$  to $-1.1$.   We have proposed a possible 
explanation  for this in \citet{DBBC08}.  This was based on the fact that
DDO~210, where the power spectrum was measured across length-scales
$100-500 \, {\rm   pc}$, had a slope  of $-2.6$  while NGC~628, a
nearly face-on galaxy where  the power spectrum was measured across
length-scales $0.8-8 \, {\rm   kpc}$ had a slope of $-1.6$. We have
interpreted the former as three dimensional (3D) turbulence
operational at small scales whereas the latter was interpreted as two
dimensional (2D) turbulence in the plane of the galactic disk. For a
nearly face-on disk galaxy we expect the transition from 2D to 3D
turbulence to be seen at a length-scale  corresponding to the scale
height of the galaxy.  Continuing with this interpretation implies
that we have also measured   3D turbulence in NGC~3741,  and 2D
turbulence in   NGC~2403, UGC~4459,  GR~8 and AND~IV.

\begin{table}
\centering
\begin{tabular}{|l|c|c|c|c|c|}
\hline
Dwarf & DDO~210 & NGC~3741 & UGC~4459 & GR~8 & AND~IV\\
\hline \hline
(1 a) $\alpha$ & $-2.6\pm0.6$&$-2.2\pm0.4$&$-1.8\pm0.6$&$-1.1\pm0.4$&$-1.3\pm0.3$\\
(2 a) range (kpc)& $0.10-0.50$&$0.15-3.75$&$0.16-1.78$&$0.1-1.5$&$0.56-6.2$\\
(3 a) $\chi^{2}/\nu$& 0.6&0.4&0.8&0.8&0.4\\
\hline \hline
Spiral & NGC~628 & NGC~2403 & NGC~1058 & NGC~1058 &\\
\hline \hline
(1 a) $\alpha$ & $-1.7\pm0.2$&$-1.8\pm0.2$&$-1.0\pm0.2$&$-2.5\pm0.6$& - \\
(2 a) range (kpc)& $1.0-10.0$&$0.5-4.4$&$1.5-10.0$&$0.6-1.5$& - \\
(3 a) $\chi^{2}/\nu$& 0.2&0.9&$0.9$&$0.5$& -\\
\hline \hline
\end{tabular}
\caption{The results for  the 5 dwarf and 3 spiral galaxies in our
  sample.  Rows 1-3 gives (1) the best fit slope $\alpha$, (2)
  length-scales over which the power law fit is valid  and (3) the
  goodness  of fit $\chi^2$ per degree of freedom.} 
\end{table}
The slope $\alpha=-1.0\pm0.2$ gives a good fit to the power spectrum
measured at the length-scales $1.5-10.0$ kpc for NGC~1058, where for
the length scales $600$ pc$-1.5$ kpc the best fit slope is
$\alpha=-2.5\pm0.6$. This, following the argument of dimensionality, is
a transition from 2D to 3D turbulence. Since the length-scale $10.0$
kpc is definitely larger than the scale height of any spiral galaxy,
we interpret the wave-length $-1.5\pm0.3$ kpc at this transition as
the scale-height of the HI disk of NGC~1058. To  our knowledge this is
the  first observational determination of the scale-height of a nearly
face on spiral galaxy through its HI power spectrum. 

We have plan to use the same estimator to estimate the power spectrum
of a large sample of spiral galaxy and to find the correlation of the
slope of the power spectrum of those galaxies with the other
measurable dynamical parameters like scale height, total HI mass
etc. of the galaxies.

\acknowledgements 
 Authors are thank full to all LFRU SOC and LOC members to give a chance to 
present our work there.

\end{document}